\documentclass[iop,apjl]{emulateapj}

\usepackage{graphicx}  
\usepackage{dcolumn}   
\usepackage{bm}        
\usepackage{amsfonts,amsmath,amssymb,mathrsfs}
\usepackage{color}
\usepackage{epsfig}
\usepackage{natbib,times}
\usepackage{url}
\usepackage{times}

\pdfoutput=1

\shorttitle{GRMHD simulations of Merging Supermassive Black Holes}
\shortauthors{Giacomazzo et al.}

\begin{document}

\title{General Relativistic Simulations of Magnetized Plasmas around
  Merging Supermassive Black Holes}

\author{Bruno {Giacomazzo}\altaffilmark{1,2,3}, John G. {Baker}\altaffilmark{3}, M. Coleman {Miller}\altaffilmark{2,4}, Christopher S. {Reynolds}\altaffilmark{2,4}, and James R. {van Meter}\altaffilmark{3}}

\altaffiltext{1}{JILA, University of Colorado and National Institute of Standards and Technology, 440 UCB, Boulder, CO 80309, USA}
\altaffiltext{2}{Department of Astronomy, University of Maryland, College Park, MD 20742, USA}
\altaffiltext{3}{Gravitational Astrophysics Laboratory, NASA Goddard Space Flight Center, Greenbelt, MD 21114, USA}
\altaffiltext{4}{Joint Space Science Institute, University of Maryland, College Park, MD 20742, USA}

\begin{abstract}
Coalescing supermassive black hole binaries are produced by the
mergers of galaxies and are the most powerful sources of gravitational
waves accessible to space-based gravitational observatories. Some such
mergers may occur in the presence of matter and magnetic fields and
hence generate an electromagnetic counterpart. In this Letter, we
present the first general relativistic simulations of magnetized
plasma around merging supermassive black holes using the general
relativistic magnetohydrodynamic code \texttt{Whisky}. By considering
different magnetic field strengths, going from non-magnetically
dominated to magnetically dominated regimes, we explore how magnetic
fields affect the dynamics of the plasma and the possible emission of
electromagnetic signals. In particular we observe a total
amplification of the magnetic field of $\sim 2$ orders of magnitude
which is driven by the accretion onto the binary and that leads to
much stronger electromagnetic signals, more than a factor of $10^4$
larger than comparable calculations done in the force-free regime
where such amplifications are not possible.
\end{abstract}


\section{Introduction}
Space-based gravitational-wave (GW) detectors, such as the planned
eLISA/NGO and SGO detectors, are expected to detect tens of
supermassive black hole (BH) mergers per year. These detections will
provide superbly precise measurements of the redshifted masses of the
holes as well as the luminosity distance to the event. However, it is
not possible to extract the redshift directly from the GWs. For this
it is necessary to look for electromagnetic signatures that would
identify the host galaxy.  The resulting combination of the redshift
with the luminosity distance would provide a powerful cosmological
probe~\citep{2003CQGra..20S..65H,2005CQGra..22S.943B,2006ApJ...637...27K,2009CQGra..26i4021A}. It
would also allow precise tests of whether GWs travel at the speed of
light, as required by general relativity. Although the merger itself
produces no electromagnetic emission, if there are significant
electromagnetic fields or mass nearby in an accretion disk then there
are various possibilities~\citep{2011CQGra..28i4021S}. For some disk
accretion rates and binary mass ratios, the binary reaches a point in
its coalescence such that further inspiral by emission of GWs occurs
more rapidly than the disk diffuses
inward~\citep{2002ApJ...567L...9A,2005ApJ...622L..93M}.  This leads
to a hole in the disk which is filled gradually after merger, leading
to a source that brightens over weeks to years depending on various
parameters~\citep{2002ApJ...567L...9A,2005ApJ...622L..93M,2010ApJ...709..774K,2010ApJ...714..404T,2010PhRvD..81b4019S}. Several
authors have discussed consequences of the recoils from asymmetric
emission of GWs during the coalescence, from prompt shocks to delayed
emission lasting millions of
years~\citep{2008ApJ...682..758S,2008ApJ...684..835S,2009PhRvD..80b4012M,2008ApJ...676L...5L,2010MNRAS.404..947C,2010PhRvD..81d4004A,2010MNRAS.401.2021R,2010arXiv1002.4185Z}. Emission
might occur in the late inspiral from effects such as enhanced
accretion, periodic Newtonian perturbations, or shearing of the disk
due to GWs~\citep{2008PhRvL.101d1101K}. Earlier precursors are also
possible, and in some cases the error volume from the GW signal may be
small enough that the host galaxy can be identified by morphology,
mass, or by the presence of an active galactic nucleus.

In the last few years there have been a number of publications
describing the evolution of gas and magnetic fields around merging
supermassive BHs. van Meter and collaborators performed test-particle
simulations of the motion of accreting gas during the last phase of
inspiral of comparable-mass supermassive
BHs~\citep{2010ApJ...711L..89V}.  These simulations suggested that
near merger a significant fraction of particles can collide with each
other at speeds approaching the speed of light, implying that a burst
of radiation might accompany the coalescence. Other works have instead
started to investigate the effect that the merging BHs would have on
surrounding gas and the possible emission of electromagnetic
signals~\citep{2009ApJ...700..859O,2010PhRvD..81h4008F,2011PhRvD..84b4024F,2010ApJ...715.1117B,2011CQGra..28i4020B,2011arXiv1101.4684B}. At
the same time there have been the first investigations of the effect
of binary black hole (BBH) mergers on electromagnetic fields in
vacuum~\citep{2009PhRvL.103h1101P,2010PhRvD..81h4007P,2010PhRvD..81f4017M}
and in a magnetically dominated
plasma~\citep{2010Sci...329..927P,2010PhRvD..82d4045P,2011arXiv1109.1177M}.

These studies have shown how magnetic and electric fields can be
distorted by the motion of the BHs and hence lead to possible
electromagnetic emission. In particular, recent studies by Palenzuela
and
{M{\"o}sta}~\citep{2010Sci...329..927P,2010PhRvD..82d4045P,2011arXiv1109.1177M}
have raised the possibility that the motion of two BHs in a
magnetically dominated plasma, i.e., in the so called force-free
regime, could generate two separate jets, one around each BH, during
the inspiral. At the time of the merger these two collimated jets
would enter in contact and form a single jet emitted from the spinning
BH formed after the merger. It is, however, still unknown how general
this scenario is and whether the emission would be detectable.

In this Letter, we present the first results from general relativistic
magnetohydrodynamic (GRMHD) simulations of magnetized plasmas around
merging supermassive black holes. By considering the evolution of
equal-mass BBH systems in plasmas with different levels of
magnetization we fill the gap between the studies of non-magnetized
gas and the results obtained in the force-free and electro-vacuum
regimes. We use a spacelike signature $(-,+,+,+)$ and will typically
use a system of units in which $c=G=M=1$, where $M$ is the total mass
of the binary. In these units $1~M$ is equivalent to $\sim 0.14~M_8$
hr and to $\sim 4.86 \times 10^{-6}~M_8$~pc, where $M_8 \equiv M/(10^8
M_{\odot})$.

\section{Numerical Methods and Initial Data}

Most of the details on the mathematical and numerical setup used for
producing the results presented here are discussed in depth elsewhere
\citep{Pollney:2007ss,Thornburg2003:AH-finding,Giacomazzo:2007ti,Giacomazzo:2009mp,2011PhRvD..83d4014G,2011arXiv1111.3344L}. In
what follows, we limit ourselves to a brief overview.

\subsection{Magnetohydrodynamics and Einstein Equations}
\label{sec:Einsten_MHD_eqs}

The evolution of the spacetime was obtained using the \texttt{Ccatie}
code, a three-dimensional finite-differencing code providing the
solution of a conformal traceless formulation of the Einstein
equations~\citep{Pollney:2007ss}, and we used the ``moving puncture''
method and gauge conditions developed
in~\cite{2006PhRvD..73l4011V}. The GRMHD equations were instead
solved using the \texttt{Whisky}
code~\citep{Giacomazzo:2007ti,2011PhRvD..83d4014G}, which adopts a
flux-conservative formulation of the equations as presented
in~\cite{Anton05} and high-resolution shock-capturing schemes. All the
results presented here have been computed using the piecewise
parabolic method, while the Harten-Lax-van Leer-Einfeldt approximate
Riemann solver has been used to compute the
fluxes~\citep{Giacomazzo:2007ti}. All the simulations were performed
using a polytropic equation of state (EOS) with a polytropic exponent
$\gamma=4/3$ and a polytropic constant $\kappa=0.2$. We used a
polytropic EOS instead of an ideal-fluid EOS because the computation
of primitive from conservative variables in highly magnetized plasmas
is much simpler and more robust for polytropes
(see~\citealt{Giacomazzo:2007ti} for details). The main results of our
work are unaffected by this choice.

In order to guarantee the divergence-free character of the MHD
equations we evolve the vector potential as described
in~\citealt{2011PhRvD..83d4014G}. When evolving the vector potential a
gauge choice needs to be made and we here use the ``algebraic
gauge''~\citep{2012PhRvD..85b4013E} which was also used in previous
GRMHD simulations with the {\tt Whisky}
code~\citep{2011PhRvD..83d4014G,2011ApJ...732L...6R}. The code has
been validated against a series of tests in special
relativity~\citep{Giacomazzo:2005jy} and in full general
relativity~\citep{Giacomazzo:2007ti}.

Since the simulations performed here consider a plasma with a total
mass negligible with respect to the mass of the two BHs, we have
decoupled the Einstein equations from the matter dynamics, i.e., the
metric variables are evolved using Einstein equations in vacuum. The
same was done in the general relativistic hydrodynamic simulations
reported in~\cite{2010PhRvD..81h4008F}. Moreover, in order to prevent
the formation of nonphysical values in the MHD quantities, we have
excised the MHD variables inside the apparent horizon of each BH.

\subsection{Adaptive Mesh Refinement}
\label{sec:AMR}
Both the Einstein and the GRMHD equations are solved using the
vertex-centered adaptive mesh-refinement (AMR) approach provided by
the \texttt{Carpet} driver~\citep{Schnetter-etal-03b}. Our rather
basic form of AMR consists of centering the highest-resolution level
around each BH and in moving the ``boxes'' following the
position of the two BHs. For the results presented here we have used
$11$ refinement levels with the finest resolution being $0.0375 M$ and
the coarsest resolution being $38.4 M$. The finest grid has a radius
of $6 M$ whereas the coarsest grid extends to $1536 M$. The large
extent of our finest grid allows us to follow accurately the dynamics
of the plasma around the BHs and it is also sufficiently large to
avoid the spurious magnetic field amplifications that may occur when
evolving the vector potential with the ``algebraic
gauge''~\citep{2012PhRvD..85b4013E}.

For all the simulations reported here we have also used a
reflection-symmetry condition across the $z=0$ plane and a
$\pi$-symmetry condition across the $x=0$ plane.\footnote{Stated
  differently, we evolve only the region $\{x\geq 0,\,z\geq 0\}$
  and apply a $180^{\circ}$-rotational-symmetry boundary condition
  across the plane at $x=0$.}

\subsection{Initial Data}
We have considered equal-mass systems of two non-spinning BHs with an
initial separation of $8.48 M$; these inspiral for approximately three
orbits before merger. The initial data were computed using the public
available \texttt{TwoPuncture} code developed
in~\cite{2004PhRvD..70f4011A} and we chose the momentum of the
punctures in order to ensure that the orbit of the two BHs is
quasicircular. We have considered two models: \texttt{B0} is
surrounded by a non-magnetized plasma, and model \texttt{B2} instead
has an initially uniform magnetic field with a ratio of magnetic to
gas pressure ($p_{\mathrm{mag}}/p_{\mathrm{gas}}\equiv\beta^{-1}$)
initially equal to $2.5 \times 10^{-2}$. The magnetic field is aligned
with the total angular momentum of the system, i.e., $B^i=(0,0,B^z)$
while in all the models the rest-mass density $\rho$ is initially
uniform and fills the entire domain; we have in mind a flow that could
be advection-dominated close to the holes (e.g.,
\citealt{1977ApJ...214..840I}) and thus have a high enough radial
velocity that it can keep up with the binary inspiral throughout the
entire coalescence. The initial distribution of the magnetic field is
similar to that adopted in previous works
\citep{2010Sci...329..927P,2010PhRvD..82d4045P,2011arXiv1109.1177M}
and it assumes that the magnetic field is anchored to a circumbinary
disk located far outside of the numerical domain. We note that these
initial data are similar to what was used in previous
non-magnetic~\citep{2010PhRvD..81h4008F} and
force-free~\citep{2010Sci...329..927P,2010PhRvD..82d4045P,2011arXiv1109.1177M}
analyses. After this Letter was submitted,
\citealt{2012arXiv1204.1073N} presented preliminary results describing
the evolution of a magnetized circumbinary accretion disk up to few
orbits before the BH merger. Their simulations showed that magnetized
plasma can indeed flow in the region where the BHs merge, and in
future simulations we plan to use initial conditions more similar to
the end result of the simulations recently reported
in~\cite{2012arXiv1204.1073N}.

 \begin{figure*}[p!]
   \begin{center}
     \begin{tabular}{cc}
       \includegraphics[width=0.4\textwidth]{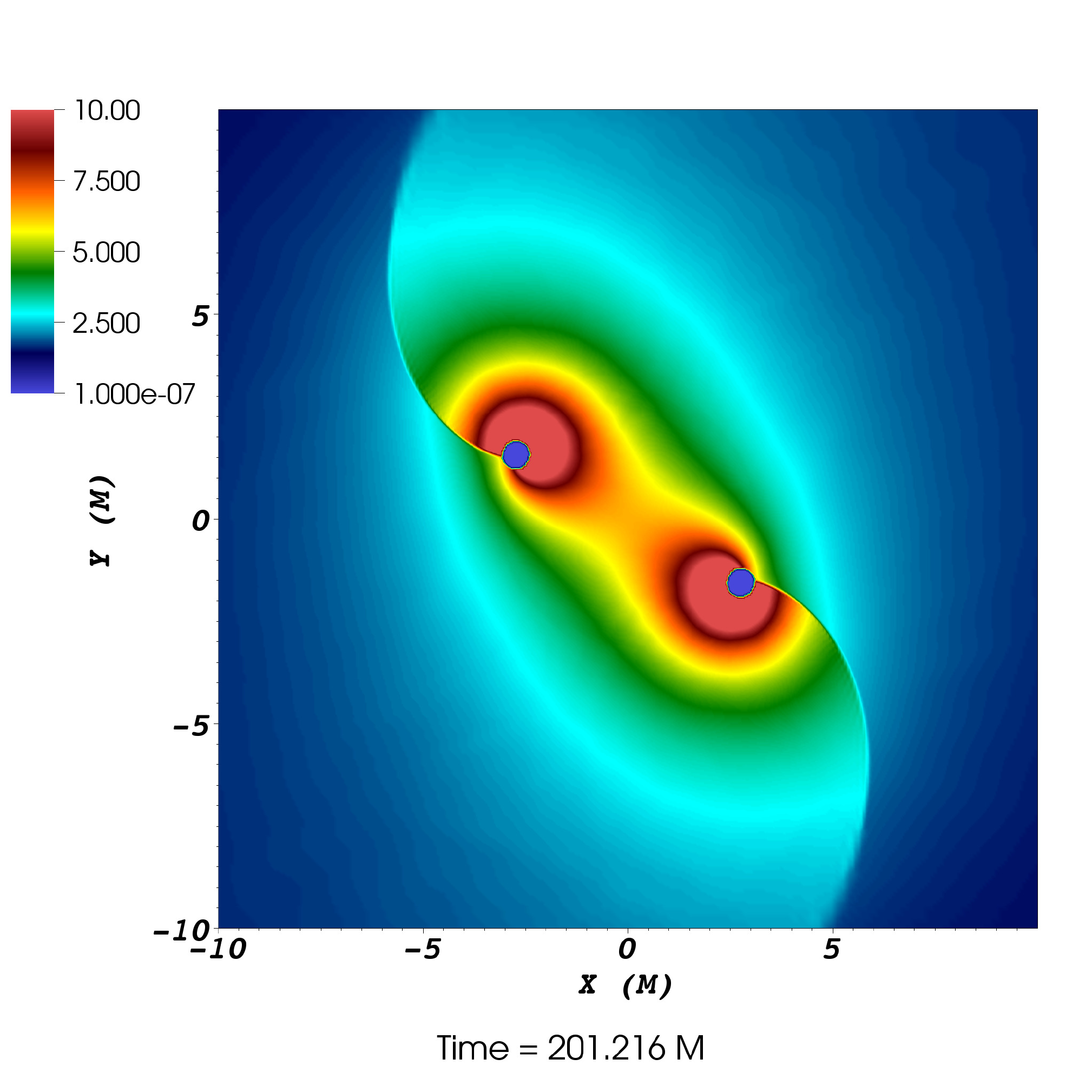}
       \includegraphics[width=0.4\textwidth]{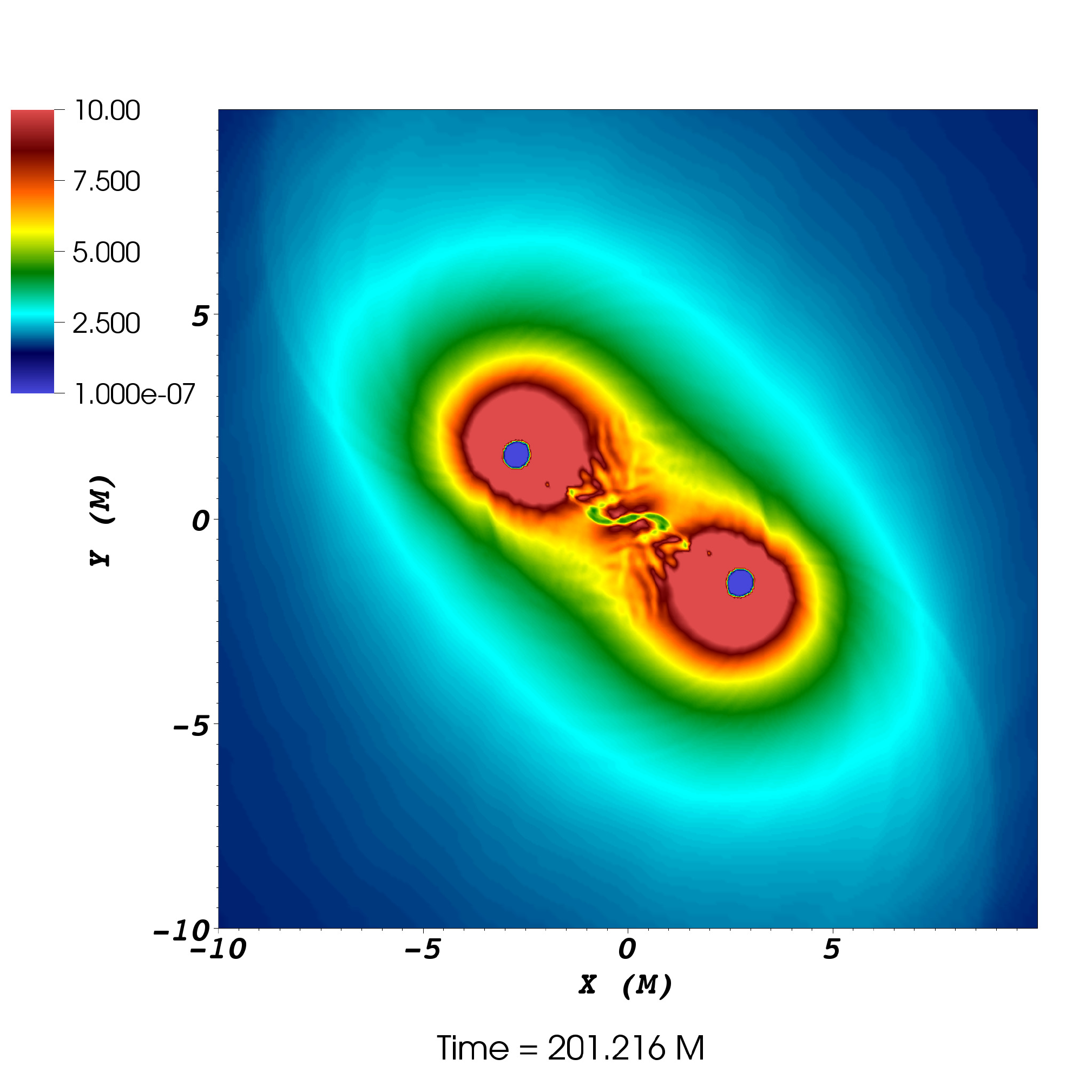}\\
       \includegraphics[width=0.4\textwidth]{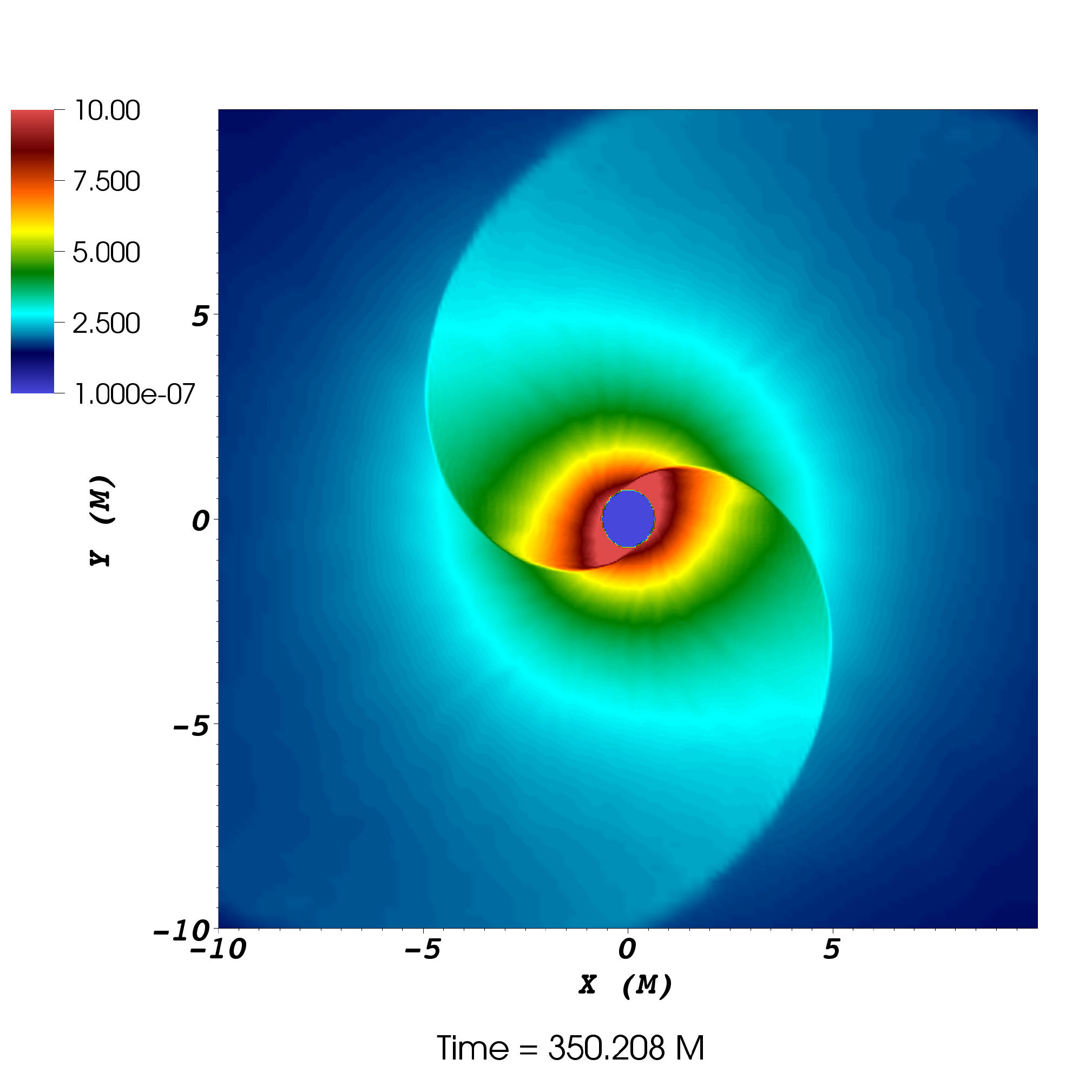}
       \includegraphics[width=0.4\textwidth]{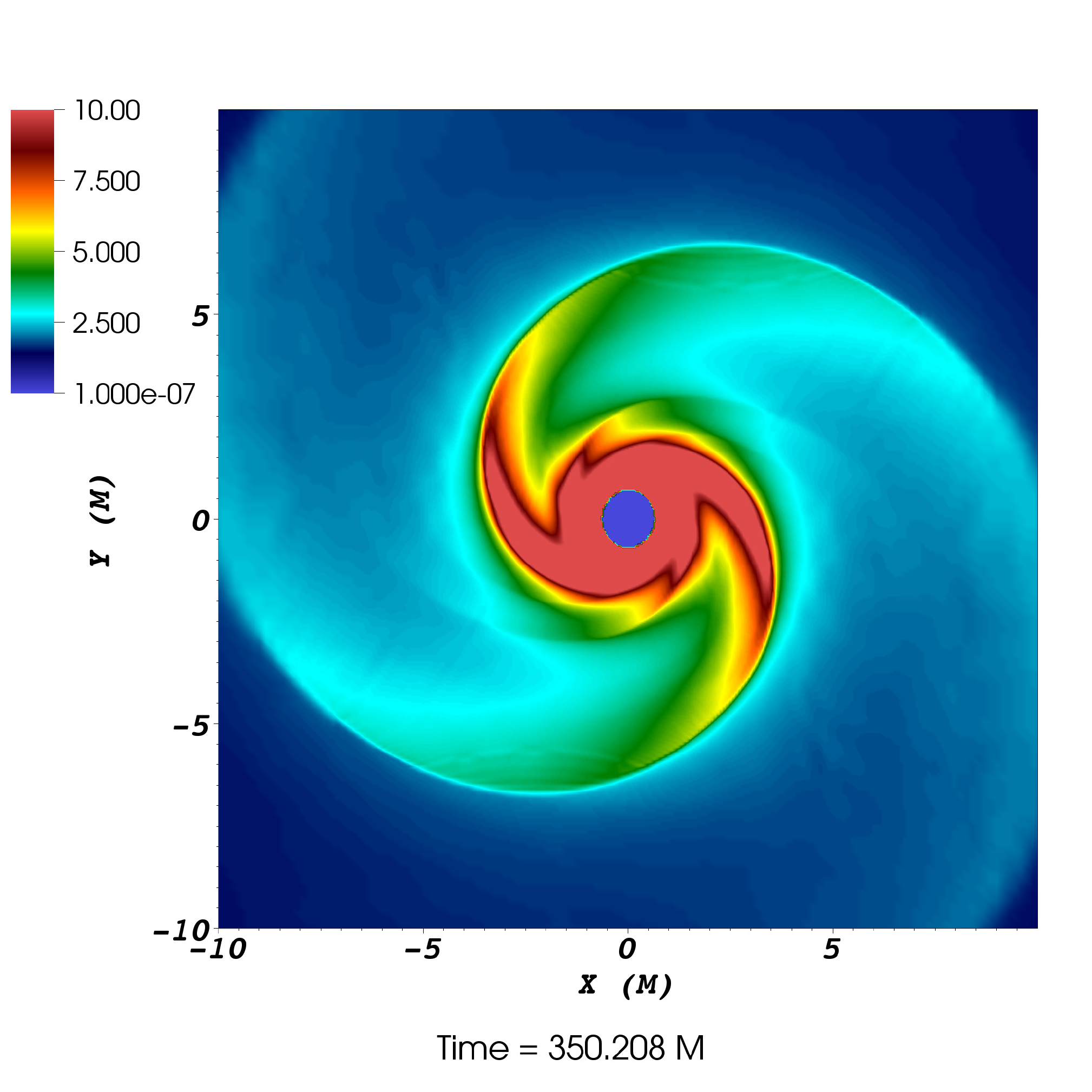}\\
       \includegraphics[width=0.4\textwidth]{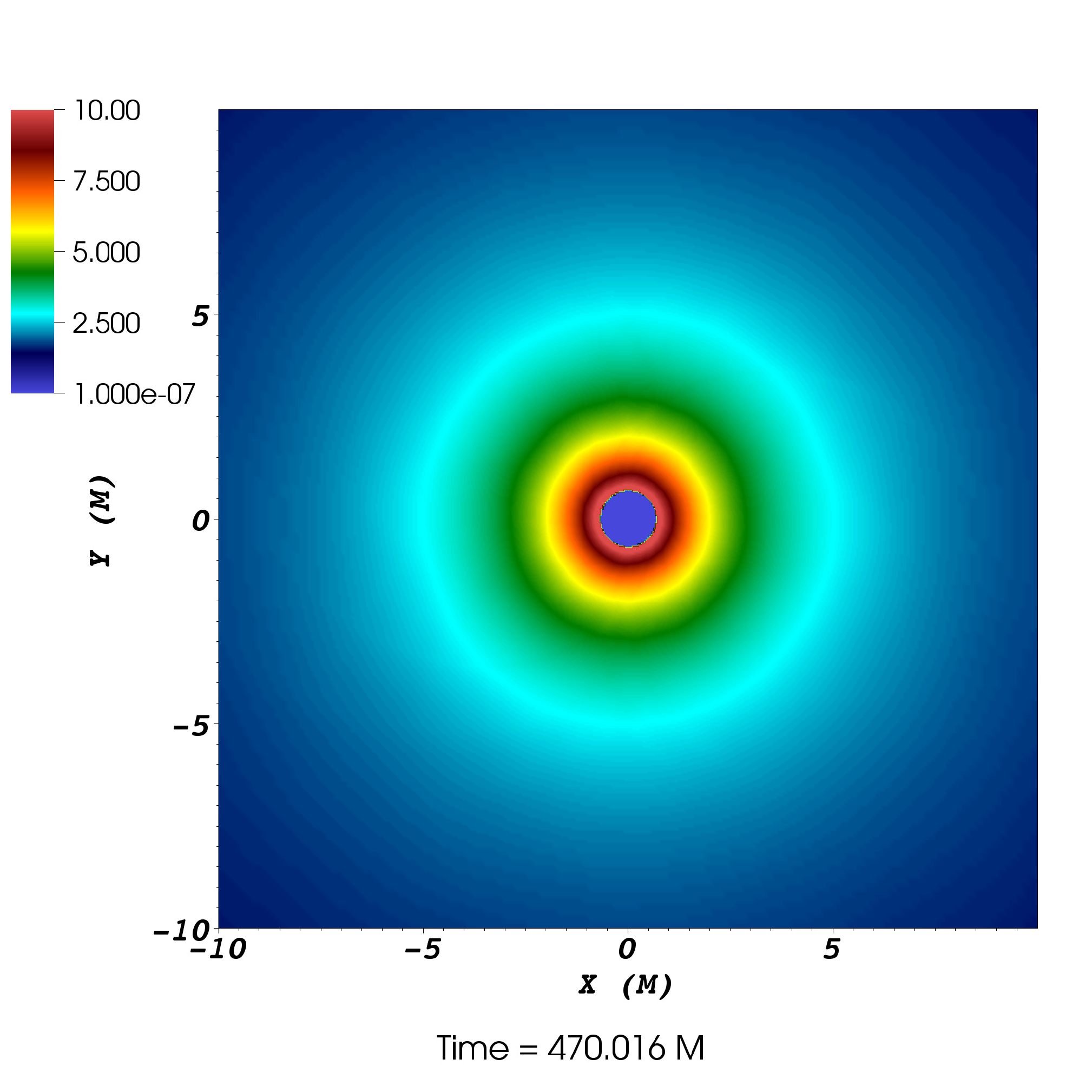}
       \includegraphics[width=0.4\textwidth]{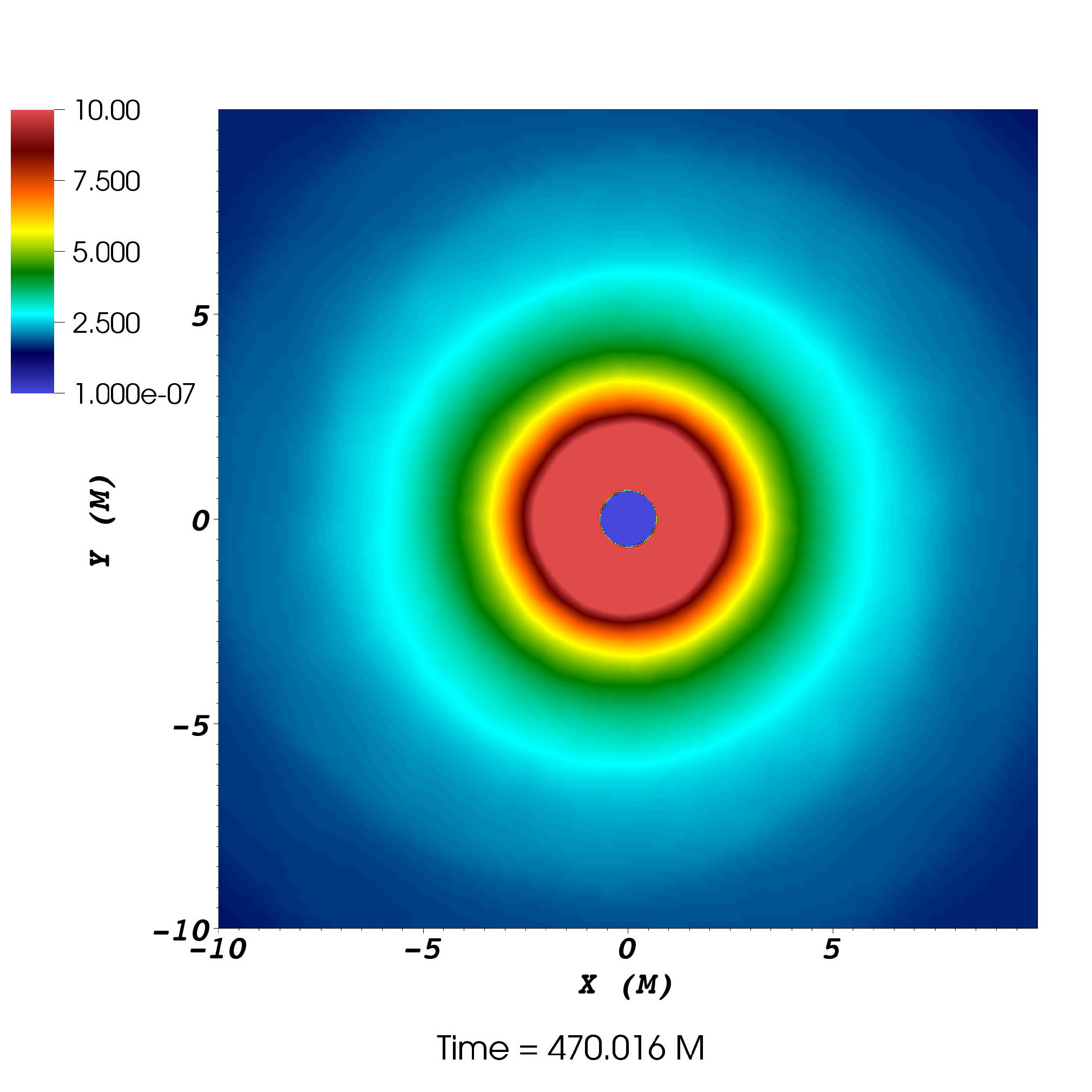}
     \end{tabular}
   \end{center}
   \caption{\label{fig:rho_xy_3orbits}Evolution of the rest-mass
     density $\rho$ on the equatorial plane for the non-magnetized
     model \texttt{B0} (left panels) and for the magnetized case
     \texttt{B2} (right panels). The units of time (shown at the
     bottom of each panel) and distance are $M$ and the rest-mass
     density is normalized to its initial value.}
 \end{figure*}

\section{Dynamics}

 \begin{figure*}[t!]
   \begin{center}
     \begin{tabular}{cc}
       \includegraphics[width=0.4\textwidth]{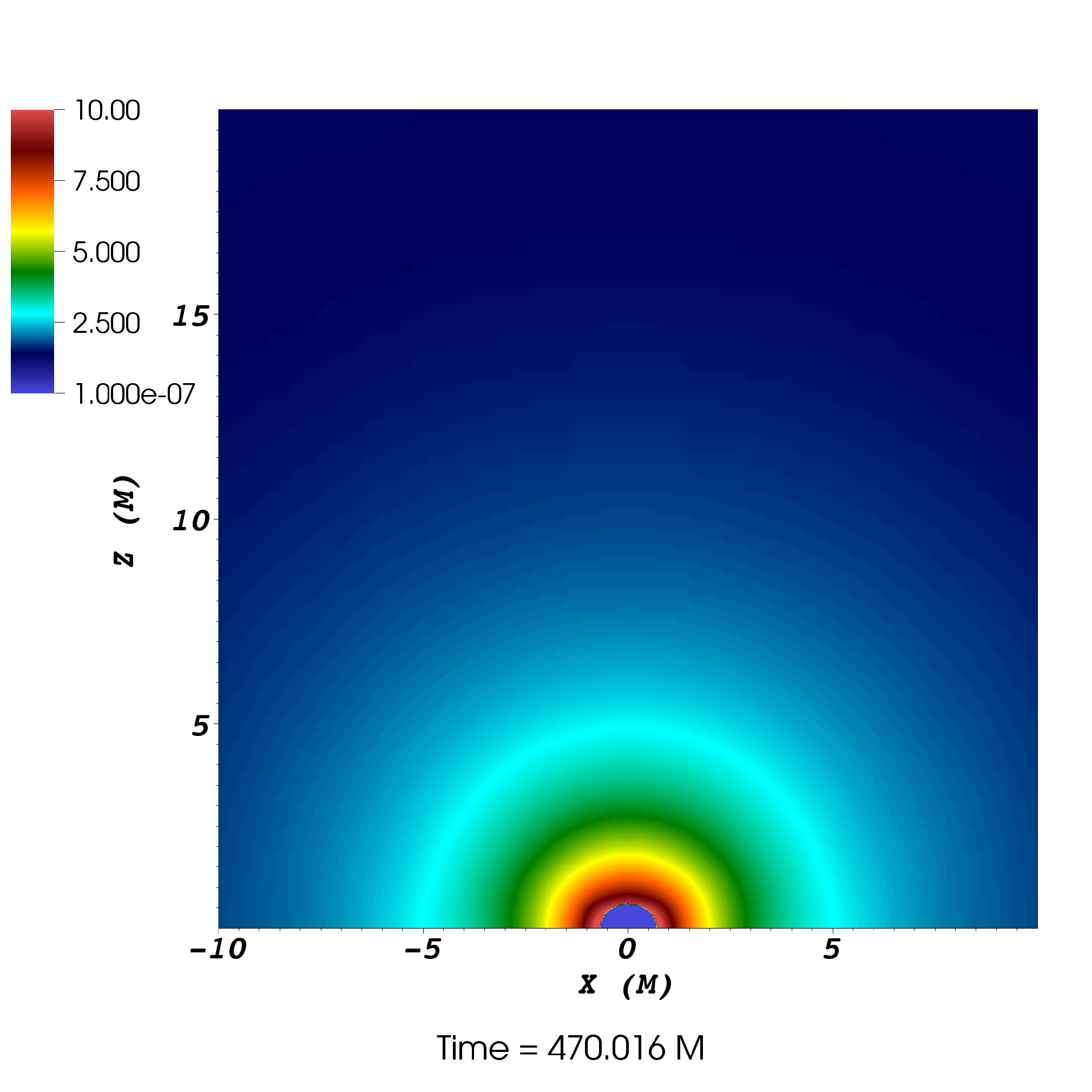}
       \includegraphics[width=0.4\textwidth]{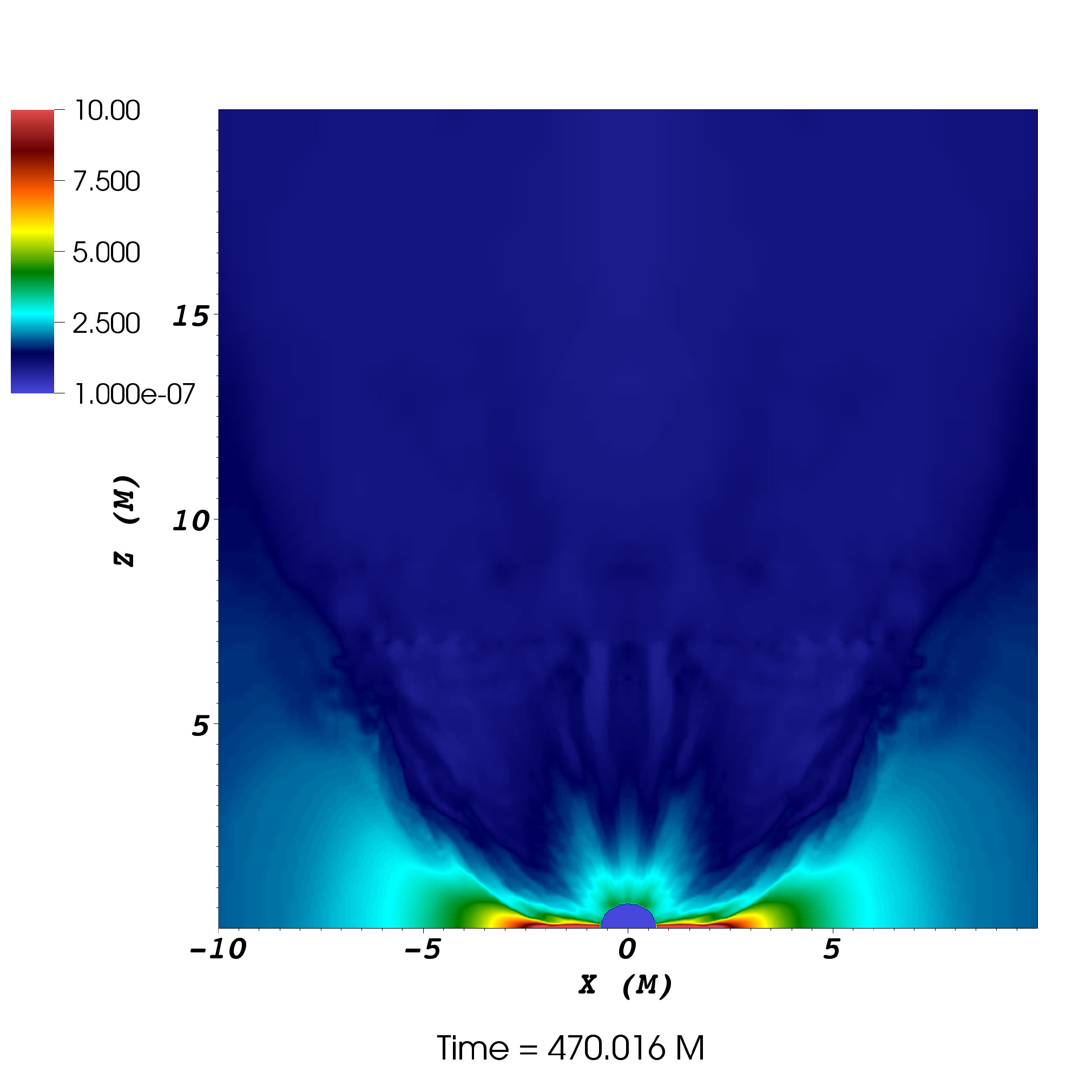}
     \end{tabular}
   \end{center}
   \caption{\label{fig:rho_xz_3orbits}Rest-mass density $\rho$ on the
     $xz$ plane for the non-magnetized model \texttt{B0} (left panel)
     and for the magnetized case \texttt{B2} (right panel) at the end
     of the simulation ($t\sim470M$). The unit of distance is $M$ and
     the rest-mass density is normalized to its initial value.}
 \end{figure*}

 \begin{figure*}[t!]
   \begin{center}
     \begin{tabular}{cc}
       \includegraphics[width=0.4\textwidth]{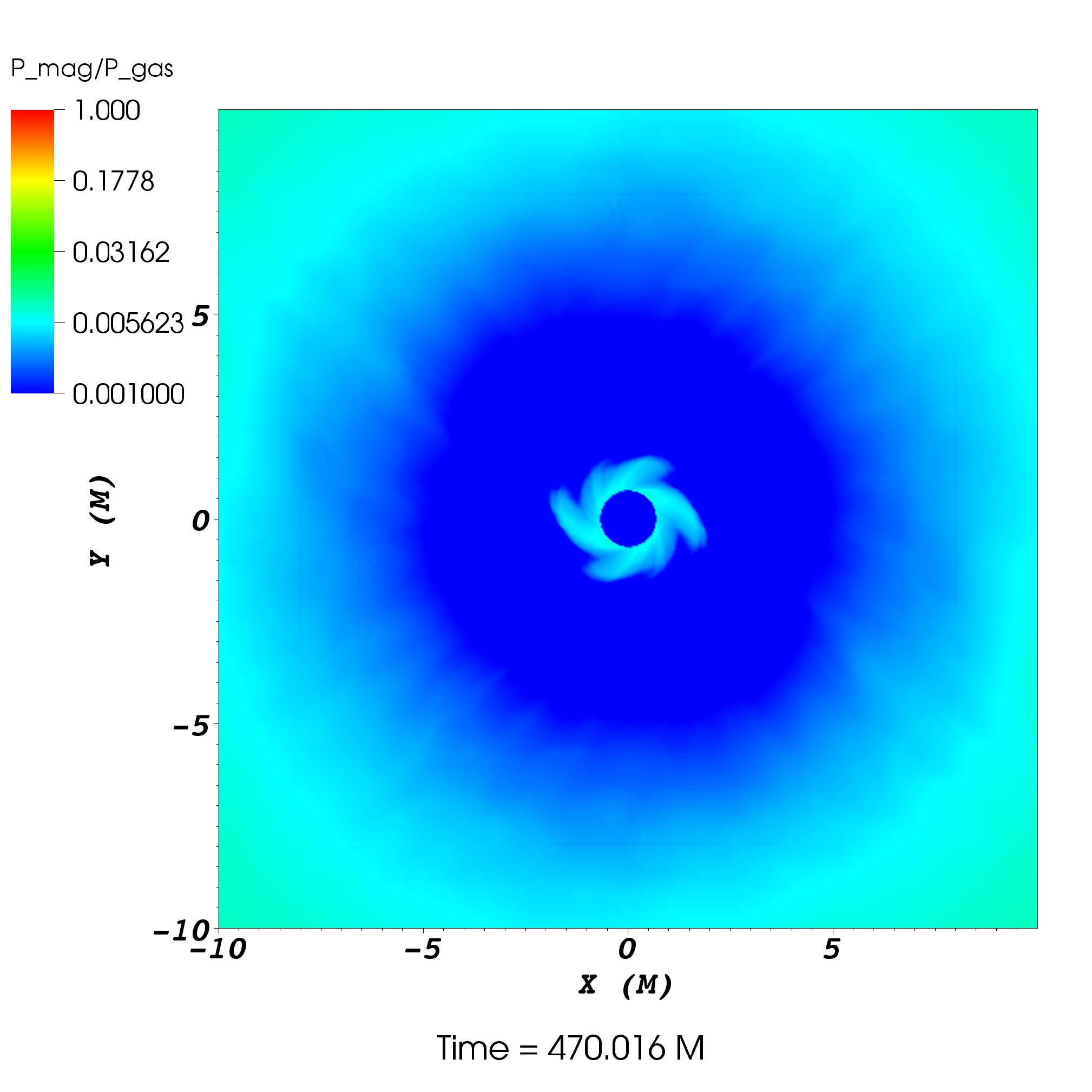}
       \includegraphics[width=0.4\textwidth]{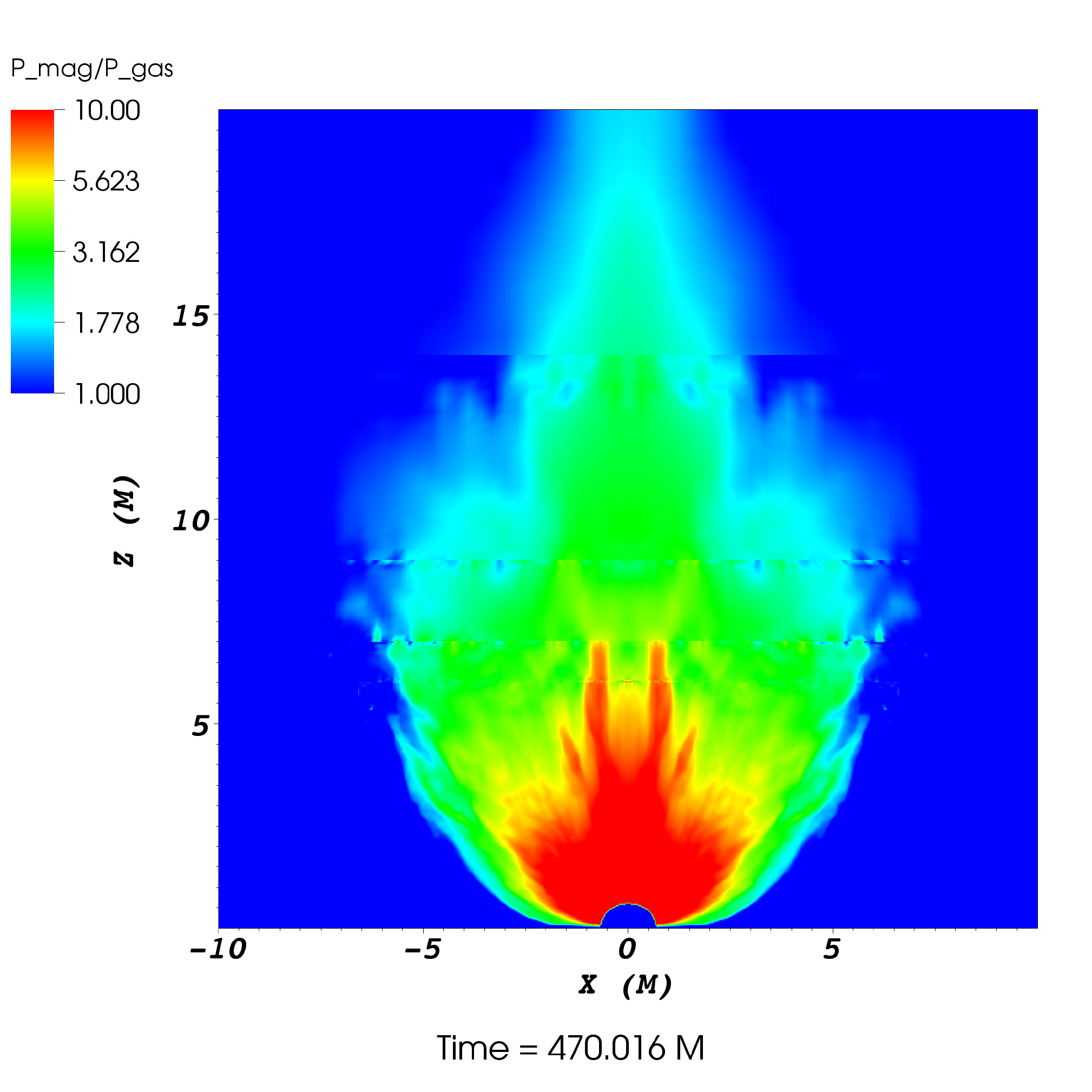}
     \end{tabular}
   \end{center}
   \caption{\label{fig:beta_3orbits}Ratio of magnetic to gas pressure
     for the magnetized model \texttt{B2} on the $xy$ (left panel) and
     $xz$ planes (right panel) at the end of the simulation
     ($t\sim470M$). The unit of distance is $M$. Note that the color
     bar is different between these two panels, on the left panel any
     magnetically dominated region would be in red while in the right
     panel the minimum value of the color bar is $p_{\mathrm{mag}}/p_{\mathrm{gas}}=1$
     in order to highlight only those regions that are magnetically
     dominated. The right panel contains five refinement levels; the
     finest refinement extends to $z=6$ and the coarsest applies to
     $z=14-20$.}
 \end{figure*}

 \begin{figure*}[t!]
   \begin{center}
     \begin{tabular}{ccc}
       \includegraphics[width=0.3\textwidth]{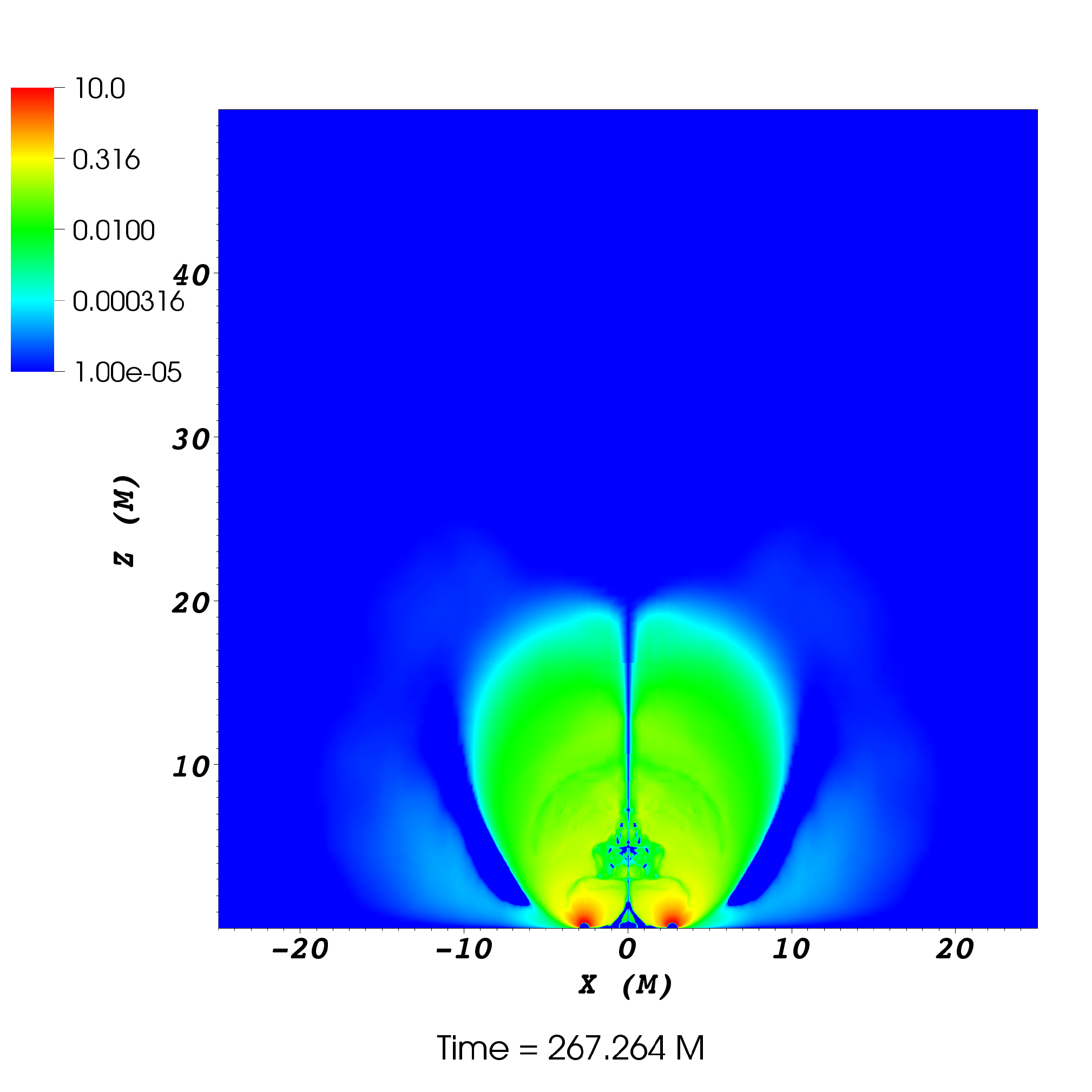}
       \includegraphics[width=0.3\textwidth]{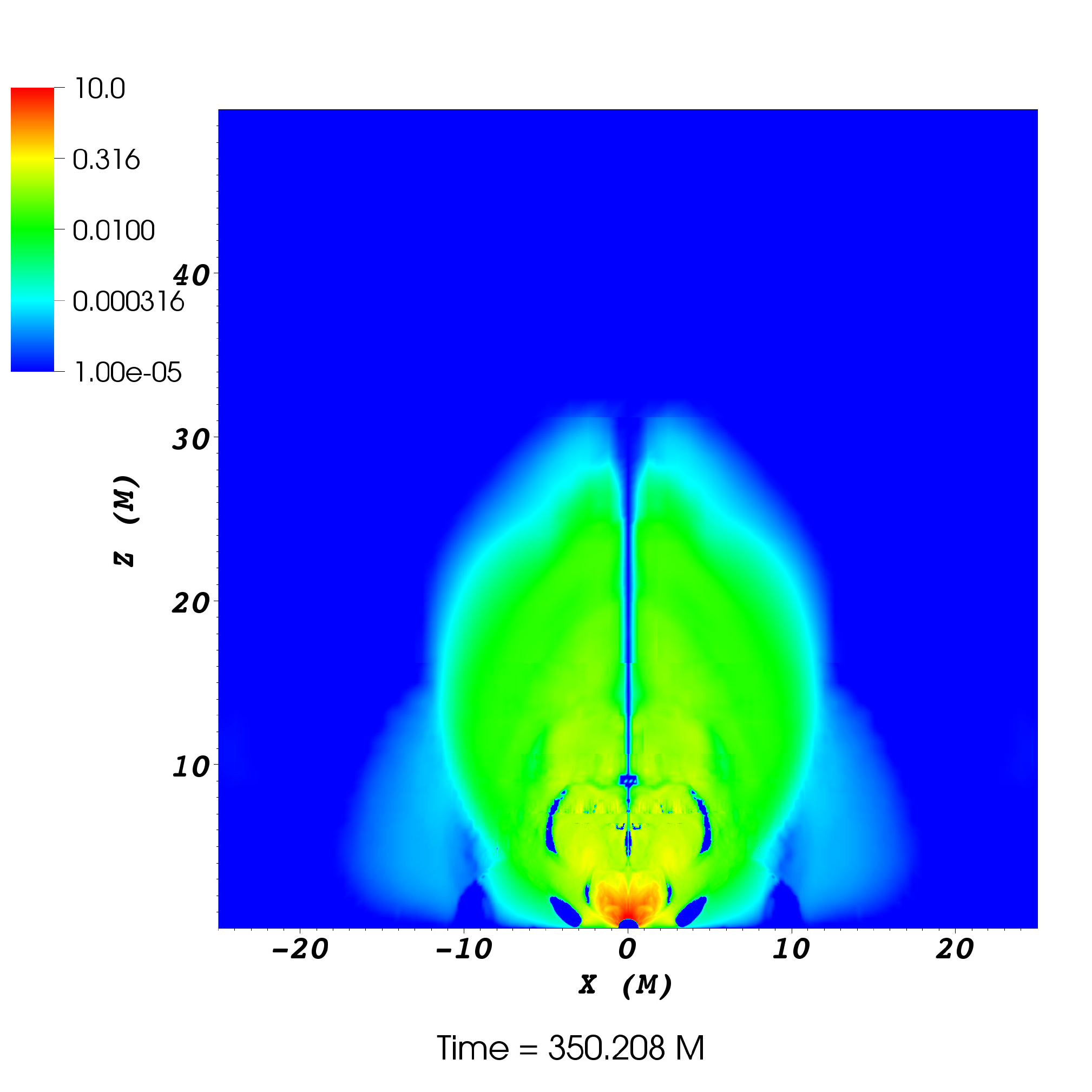}
       \includegraphics[width=0.3\textwidth]{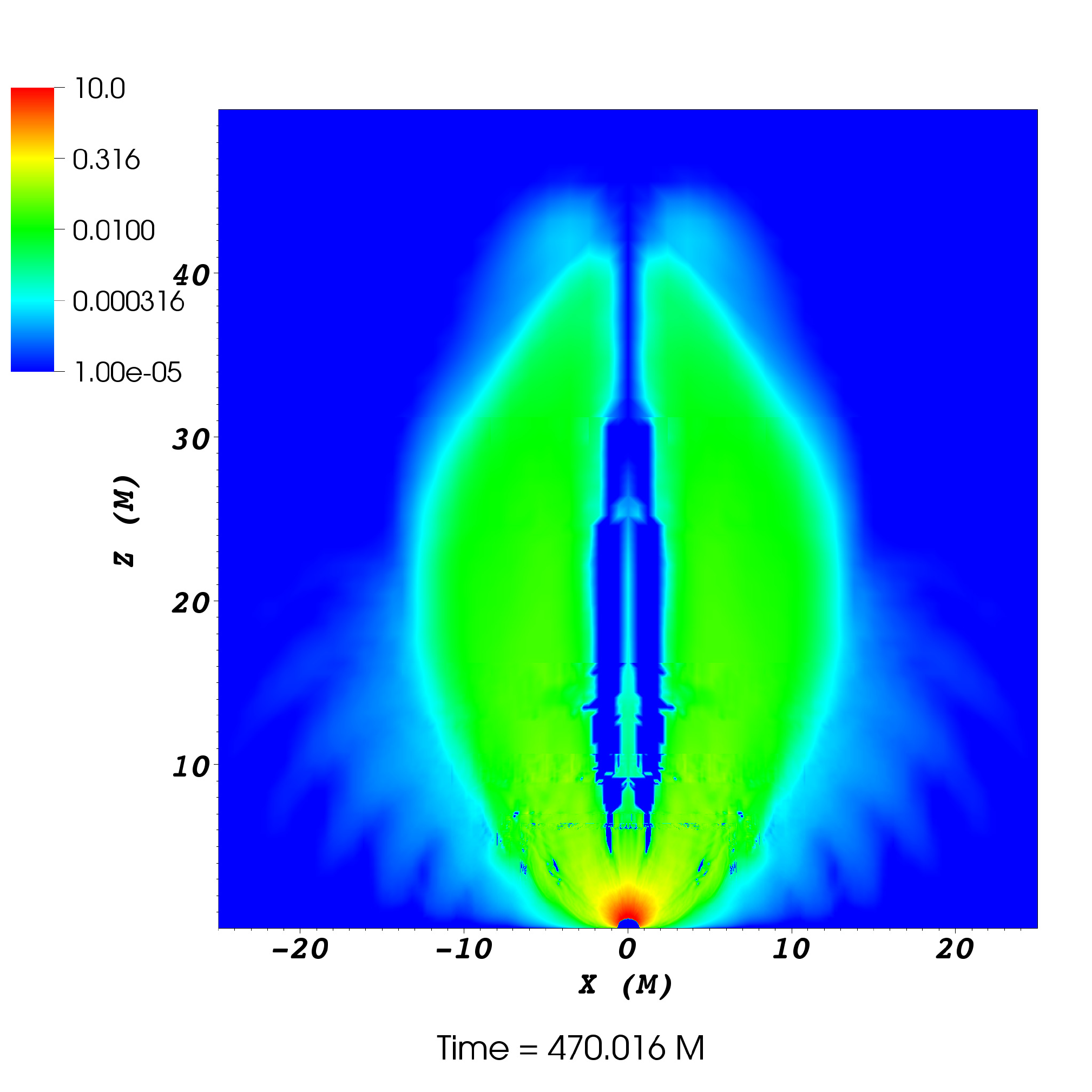}
     \end{tabular}
   \end{center}
   \caption{\label{fig:Sz_xz_3orbits} Evolution of the Poynting vector
     on the $xz$ plane for the magnetized model B2. The panels refer
     respectively to the time after two orbits, at the merger and at
     the end of the simulation. The units of time (shown at the bottom
     of each panel) and distance are $M$.}
 \end{figure*}

Figure~\ref{fig:rho_xy_3orbits} shows the evolution of the rest-mass
density $\rho$ on the equatorial plane for models \texttt{B0} (left
panels) and \texttt{B2} (right panels). The evolution of the
non-magnetized model (left panels) resembles the one described
in~\cite{2010PhRvD..81h4008F} with the formation of two spiral shocks
during the inspiral and the formation of a central spinning BH
surrounded by a spherical distribution of matter accreting onto
it. The magnetized model (right panels) shows quite different
dynamics. During the evolution the magnetic field strength increases
by approximately two orders of magnitude and contributes significantly
to the total pressure in the gas. Because of
this,\footnote{See~\cite{2009ApJ...690L..47M} for an example of how the
  magnetic pressure can affect shock formation.} the two shock waves
that are present during the inspiral of model \texttt{B0} are strongly
reduced and hardly visible (first panel in the right
column). Moreover, the density close to each of the two BHs and in the
region connecting the two BHs is much larger than in the unmagnetized
case. In this region it is also possible to see the formation of
instabilities that are not present without magnetic fields. Soon after
the merger (second panel in the right column) the spinning BH is
surrounded by a disk with a density a factor of $\sim 3$ larger than
in the unmagnetized case, and two shock waves are formed and the
system finally relaxes to its final configuration (third panel in the
right column).  The temperature in the magnetic simulation is larger
by up to $\sim 40 \%$ for model \texttt{B2} than in the unmagnetized
model.

Although the evolution on the equatorial plane already shows some
differences due to the effect of the magnetic field, the main
difference is in the rest-mass density on the
meridional plane. In Figure~\ref{fig:rho_xz_3orbits} we show the
rest-mass density $\rho$ on the $xz$ plane for models \texttt{B0} (left
panel) and \texttt{B2} (right panel) at the end of the simulation
($t\sim470M$). Whereas the end result of the evolution of the unmagnetized
model is a plasma accreting spherically onto a spinning BH, in the case
of the magnetized model \texttt{B2} the system forms a thin accretion
disk and a funnel is created around the spin axis of the BH. Although at
this time no relativistic jet is emitted (the Lorentz factor is lower
than $\sim 3$ at the end of the simulation), such emission might exist
at later times (which are outside the scope of the present Letter).

The difference between the rest-mass density distribution in the
unmagnetized and magnetized cases can be better understood by looking
at Figure~\ref{fig:beta_3orbits}, which shows the ratio of magnetic to
gas pressure (i.e., the inverse of the plasma parameter $\beta$) for
the magnetized model \texttt{B2} on the $xy$ (left panel) and $xz$
planes (right panel) at the end of the simulation ($t\sim470M$).  On
the equatorial plane no region is magnetically dominated and the
central region inside the disk has larger values of $\beta$ than in
the initial conditions, but the $xz$ plane shows clearly the presence
of a strongly magnetically dominated region close to the spin axis of
the BH with $\beta\sim 10^{-2}$. During the inspiral and merger
magnetic field lines are indeed compressed and twisted causing the
magnetic field to be amplified of approximately two orders of
magnitude. This highly magnetized region is responsible for the
creation of the thin disk and its funnel. If these simulations had
been run for many orbital times (which would have required much larger
computational resources), the amplification of the magnetic field
would likely have been even larger, due to effects like the
magneto-rotational instability~\citep{1991ApJ...376..214B}.

\section{Electromagnetic emission}
In Figure~\ref{fig:Sz_xz_3orbits} we show the evolution of the $z$
component of the Poynting vector on the $xz$ plane for model
\texttt{B2}. We show in particular its outgoing component after two
orbits (first panel), at the time of the merger (second panel) and at
the end of the simulation (third panel). One of the main differences
with respect to force-free simulations is that we do not observe the
two strong and distinct jets originating from each BH and the ``jet''
that is emitted by the system propagates slowly into the medium
surrounding the binary. Whereas in the force-free scenarios the jet
propagates in a very low-density medium where the inertia of the
plasma can be neglected, in our simulations the jet has to break
through the infalling medium outside the binary. We also note that in
our scenario the emission is mainly collimated and parallel to the
angular momentum of the binary and to the spin of the final BH with no
sign of the dominant quadrupolar emission that was observed in the
recent force-free simulations of~\cite{2011arXiv1109.1177M}. We
also note that the non-collimated emission discovered
in~\cite{2011arXiv1109.1177M} is larger than their collimated
emission and it is $\sim 600$ times smaller than our luminosity.

\begin{figure}[t!]
  \begin{center}
    \includegraphics[width=0.39\textwidth]{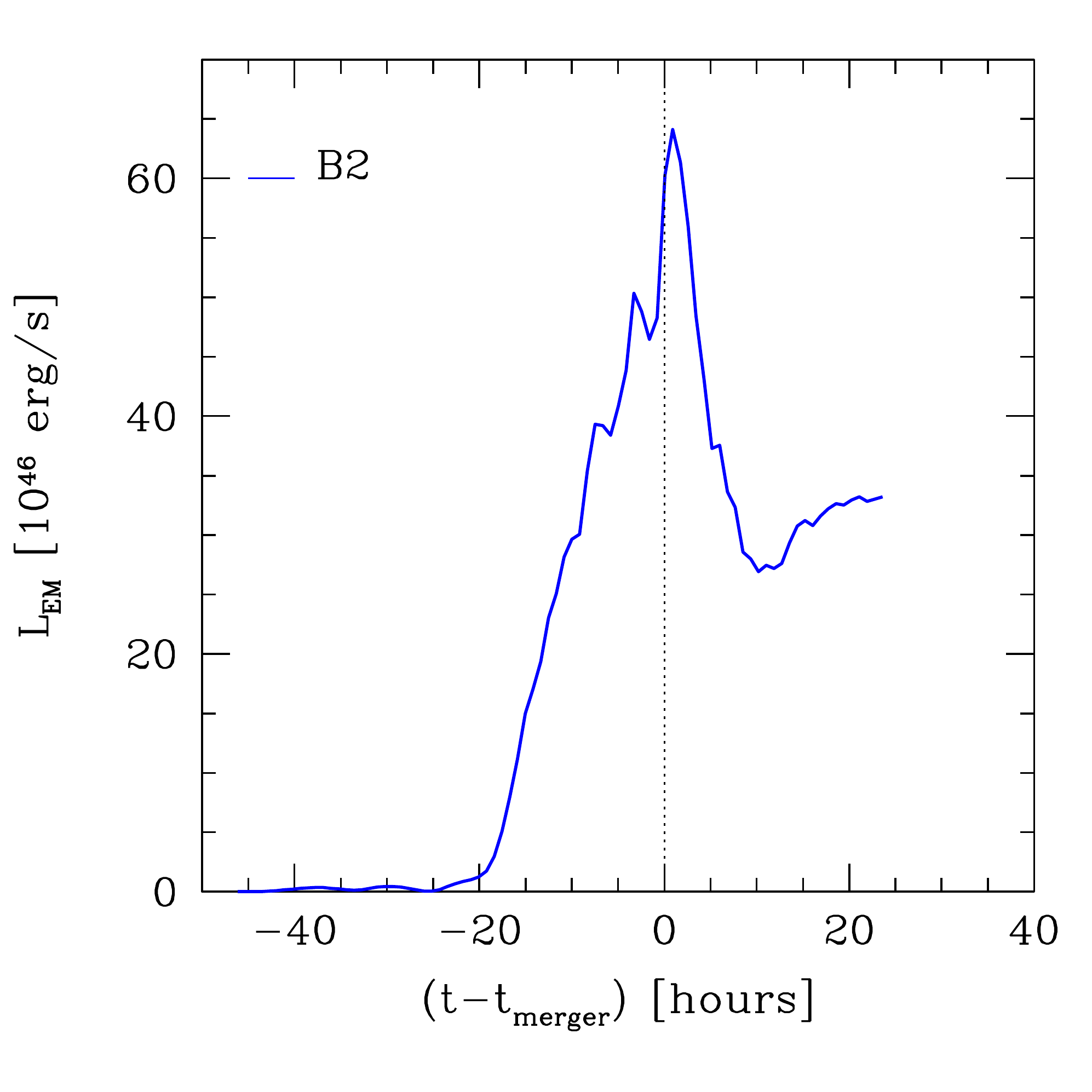}
  \end{center}
  \caption{\label{fig:Lum}Evolution of the luminosity for the
    magnetized model \texttt{B2} (blue solid line). The luminosity is
    computed at a distance $z=10M$ for a binary system with a total
    mass of $10^8 M_{\odot}$, an initial rest-mass density of
    $10^{-11} \mathrm{g~cm^{-3}}$, and an initial magnetic field of
    $\sim 10^4 \mathrm{G}$.}
\end{figure}

Finally, in Figure~\ref{fig:Lum} we show the Poynting flux luminosity
computed for model \texttt{B2} (blue solid line). We have rescaled our
results to consider a binary system with a total mass of $10^8
M_{\odot}$ and immersed in a plasma with a rest-mass density
$\rho=10^{-11}\mathrm{g\, cm^{-3}}$. This corresponds to having an
initial magnetic field of $\sim 10^4 \mathrm{G}$ for model
\texttt{B2}, which is also the same magnetic field strength considered
in~\cite{2010Sci...329..927P}. The luminosity is computed at a
distance of $z=10M$. Model \texttt{B2} shows the characteristic
increase in luminosity during the inspiral, with a peak corresponding
to the time of the merger, followed by a drop-off of a factor $\sim
2$.  This is qualitatively similar to what is observed in force-free
simulations~\citep{2010Sci...329..927P}, but our actual luminosities
are considerably higher. This happens because in our ideal GRMHD
simulations the magnetic field is amplified of $\sim 2$ orders of
magnitude. So even when starting with a magnetic field of $\sim 10^4
\mathrm{G}$, the final configuration has a field of $\sim 10^6
\mathrm{G}$. If we were to compare with force-free simulations
starting with such a high field we would obtain similar
luminosities~\citep{2010Sci...329..927P}. Indeed, we note that our
luminosity at the end is $\sim 10^{47} \mathrm{erg\, s^{-1}}$ and in
the simulations reported in~\cite{2010Sci...329..927P} is $\sim
10^{43} \mathrm{erg\, s^{-1}}$ for a magnetic field that is $\sim 2$
orders of magnitude lower than the one we have at the end of our
simulations. Since the luminosity $L_{\mathrm{EM}}$ scales as $B^2$ our values
are consistent with those one would observe in a force-free
regime. This again highlights one of the main differences between our
simulations and those assuming a force-free regime since the
beginning. Because of the accretion of the plasma onto the BHs,
magnetic field lines are compressed and twisted driving the large
amplification we observe. In a force-free regime, the magnetic field
is decoupled from the dynamics of the matter and such large
amplifications cannot be obtained.

\section{\label{sec:conclusions}Conclusions}
We have presented the first numerical GRMHD simulations of magnetized
plasmas around merging supermassive BBHs. We have for the first time
investigated the role of magnetic fields in the plasma dynamics and
filled the gap between the works that have considered non-magnetized
gas and the results obtained in the force-free and electro-vacuum
regimes.

We have shown that even plasmas that are initially not magnetically
dominated have different dynamics than in the unmagnetized case and
that magnetic plasmas can generate strong and collimated
electromagnetic emission.  We therefore generalize the physical
regimes of matter and electromagnetic fields around coalescing BHs
that can lead to potentially detectable emission.

\acknowledgments We thank Phil Armitage, Tamara Bogdanovic, Bernard
Kelly, Krzysztof Nalewajko, Carlos Palenzuela, Luciano Rezzolla,
Jeremy Schnittman, and Roman Shcherbakov for useful comments and
suggestions. We also thank Philip Cowperthwaite for help in
visualizing some of the numerical data. Resources supporting this work
were provided by the NASA High-End Computing (HEC) program through the
NASA Advanced Supercomputing (NAS) Division at Ames Research Center
and NASA Center for Climate Simulation (NCCS) at Goddard Space Flight
Center. Numerical simulations were also performed on the cluster
RANGER at the Texas Advanced Computing Center (TACC) at The University
of Texas at Austin through XSEDE grant no. TG-PHY110027. B.G.
acknowledges support from NASA grant no. NNX09AI75G and NSF grant
no. AST 1009396. J.B. and J.v.M. acknowledge support from NASA grant
09-ATP09-0136.

\bibliographystyle{apj}

\begin{thebibliography}{49}
\expandafter\ifx\csname natexlab\endcsname\relax\def\natexlab#1{#1}\fi

\bibitem[{{Anderson} {et~al.}(2010){Anderson}, {Lehner}, {Megevand}, \&
  {Neilsen}}]{2010PhRvD..81d4004A}
{Anderson}, M., {Lehner}, L., {Megevand}, M., \& {Neilsen}, D. 2010, \prd, 81,
  044004

\bibitem[Ansorg et al.(2004)]{2004PhRvD..70f4011A} Ansorg, M.,
  Br{\"u}gmann, B., \& Tichy, W.\ 2004, \prd, 70, 064011

\bibitem[{Ant{\'o}n {et~al.}(2006)Ant{\'o}n, Zanotti, Miralles, Mart{\'\i},
  Ib{\'a}{\~n}ez, Font, \& Pons}]{Anton05}
Ant{\'o}n, L., Zanotti, O., Miralles, J.~A., {et~al.} 2006, \apj, 637,
  296

\bibitem[Armitage \& Natarajan(2002)]{2002ApJ...567L...9A} Armitage,
  P.~J., \& Natarajan, P.\ 2002, \apjl, 567, L9

\bibitem[{{Arun} {et~al.}(2009){Arun}, {Mishra}, {Van Den Broeck}, {Iyer},
  {Sathyaprakash}, \& {Sinha}}]{2009CQGra..26i4021A}
{Arun}, K.~G., {Mishra}, C.~K., {Van Den Broeck}, C., {et~al.} 2009, Class.
  Quantum Grav., 26, 094021

\bibitem[Balbus \& Hawley(1991)]{1991ApJ...376..214B} Balbus, S.~A.,
  \& Hawley, J.~F.\ 1991, \apj, 376, 214

\bibitem[{{Berti} {et~al.}(2005){Berti}, {Buonanno}, \&
    {Will}}]{2005CQGra..22S.943B} {Berti}, E., {Buonanno}, A., \&
  {Will}, C.~M. 2005, Class. Quantum Grav., 22, 943

\bibitem[{{Bode} {et~al.}(2012){Bode}, {Bogdanovic}, {Haas}, {Healy}, {Laguna},
  \& {Shoemaker}}]{2011arXiv1101.4684B}
{Bode}, T., {Bogdanovic}, T., {Haas}, R., {et~al.} 2012, \apj, 744, 45 

\bibitem[{{Bode} {et~al.}(2010){Bode}, {Haas}, {Bogdanovi{\'c}}, {Laguna}, \&
  {Shoemaker}}]{2010ApJ...715.1117B}
{Bode}, T., {Haas}, R., {Bogdanovi{\'c}}, T., {Laguna}, P., \& {Shoemaker}, D.
  2010, \apj, 715, 1117

\bibitem[{{Bogdanovi{\'c}} {et~al.}(2011){Bogdanovi{\'c}}, {Bode}, {Haas},
  {Laguna}, \& {Shoemaker}}]{2011CQGra..28i4020B}
{Bogdanovi{\'c}}, T., {Bode}, T., {Haas}, R., {Laguna}, P., \& {Shoemaker}, D.
  2011, Class. Quantum Grav., 28, 094020

\bibitem[{{Corrales} {et~al.}(2010){Corrales}, {Haiman}, \&
  {MacFadyen}}]{2010MNRAS.404..947C}
{Corrales}, L.~R., {Haiman}, Z., \& {MacFadyen}, A. 2010, \mnras, 404, 947

\bibitem[Etienne et al.(2012)]{2012PhRvD..85b4013E} Etienne, Z.~B., 
Paschalidis, V., Liu, Y.~T., \& Shapiro, S.~L.\ 2012, \prd, 85, 024013 

\bibitem[{{Farris} {et~al.}(2010){Farris}, {Liu}, \&
  {Shapiro}}]{2010PhRvD..81h4008F}
{Farris}, B.~D., {Liu}, Y.~T., \& {Shapiro}, S.~L. 2010, \prd, 81, 084008

\bibitem[Farris et al.(2011)]{2011PhRvD..84b4024F} Farris, B.~D., Liu, 
Y.~T., \& Shapiro, S.~L.\ 2011, \prd, 84, 024024 

\bibitem[{Giacomazzo \& Rezzolla(2006)}]{Giacomazzo:2005jy}
Giacomazzo, B., \& Rezzolla, L. 2006, J. Fluid Mech., 562, 223

\bibitem[{Giacomazzo \& Rezzolla(2007)}]{Giacomazzo:2007ti}
Giacomazzo, B., \& Rezzolla, L. 2007, Class. Quantum Grav., 24, S235

\bibitem[{{Giacomazzo} {et~al.}(2009){Giacomazzo}, {Rezzolla}, \&
  {Baiotti}}]{Giacomazzo:2009mp}
{Giacomazzo}, B., {Rezzolla}, L., \& {Baiotti}, L. 2009, \mnras, 399, L164

\bibitem[{{Giacomazzo} {et~al.}(2011){Giacomazzo}, {Rezzolla}, \&
  {Baiotti}}]{2011PhRvD..83d4014G}
{Giacomazzo}, B., {Rezzolla}, L., \& {Baiotti}, L. 2011, \prd, 83, 044014

\bibitem[{{Hughes} \& {Holz}(2003)}]{2003CQGra..20S..65H}
{Hughes}, S.~A., \& {Holz}, D.~E. 2003, Class. Quantum Grav., 20, 65

\bibitem[{{Ichimaru}(1977)}]{1977ApJ...214..840I}
{Ichimaru}, S. 1977, \apj, 214, 840

\bibitem[{{Kocsis} {et~al.}(2006){Kocsis}, {Frei}, {Haiman}, \&
  {Menou}}]{2006ApJ...637...27K}
{Kocsis}, B., {Frei}, Z., {Haiman}, Z., \& {Menou}, K. 2006, \apj, 637, 27

\bibitem[{{Kocsis} \& {Loeb}(2008)}]{2008PhRvL.101d1101K}
{Kocsis}, B., \& {Loeb}, A. 2008, \prl, 101, 041101

\bibitem[{{Krolik}(2010)}]{2010ApJ...709..774K}
{Krolik}, J.~H. 2010, \apj, 709, 774

\bibitem[{{Lippai} {et~al.}(2008){Lippai}, {Frei}, \&
  {Haiman}}]{2008ApJ...676L...5L}
{Lippai}, Z., {Frei}, Z., \& {Haiman}, Z. 2008, \apjl, 676, L5

\bibitem[L{\"o}ffler et al.(2012)]{2011arXiv1111.3344L} L{\"o}ffler, F., 
Faber, J., Bentivegna, E., et al.\ 2012, Class. Quantum Grav., 29, 115001

\bibitem[{{Megevand} {et~al.}(2009){Megevand}, {Anderson}, {Frank},
  {Hirschmann}, {Lehner}, {Liebling}, {Motl}, \&
  {Neilsen}}]{2009PhRvD..80b4012M}
{Megevand}, M., {Anderson}, M., {Frank}, J., {et~al.} 2009, \prd, 80, 024012

\bibitem[{{Milosavljevi{\'c}} \& {Phinney}(2005)}]{2005ApJ...622L..93M}
{Milosavljevi{\'c}}, M., \& {Phinney}, E.~S. 2005, \apjl, 622, L93

\bibitem[{{Mizuno} {et~al.}(2009){Mizuno}, {Zhang}, {Giacomazzo}, {Nishikawa},
  {Hardee}, {Nagataki}, \& {Hartmann}}]{2009ApJ...690L..47M}
{Mizuno}, Y., {Zhang}, B., {Giacomazzo}, B., {et~al.} 2009, \apjl, 690, L47

\bibitem[{M{\"o}sta} et al.(2012)]{2011arXiv1109.1177M} {M{\"o}sta},
  P., Alic, D., Rezzolla, L., Zanotti, O., \& Palenzuela, C.\ 2012,
  \apjl, 749, L32

\bibitem[{{M{\"o}sta} {et~al.}(2010){M{\"o}sta}, {Palenzuela}, {Rezzolla},
  {Lehner}, {Yoshida}, \& {Pollney}}]{2010PhRvD..81f4017M}
{M{\"o}sta}, P., {Palenzuela}, C., {Rezzolla}, L., {et~al.} 2010, \prd, 81,
  064017

\bibitem[Noble et al.(2012)]{2012arXiv1204.1073N} Noble, S.~C.,
  Mundim, B.~C., Nakano, H., et al.\ 2012, arXiv:1204.1073

\bibitem[{{O'Neill} {et~al.}(2009){O'Neill}, {Miller}, {Bogdanovi{\'c}},
  {Reynolds}, \& {Schnittman}}]{2009ApJ...700..859O}
{O'Neill}, S.~M., {Miller}, M.~C., {Bogdanovi{\'c}}, T., {Reynolds}, C.~S., \&
  {Schnittman}, J.~D. 2009, \apj, 700, 859

\bibitem[{{Palenzuela} {et~al.}(2009){Palenzuela}, {Anderson}, {Lehner},
  {Liebling}, \& {Neilsen}}]{2009PhRvL.103h1101P}
{Palenzuela}, C., {Anderson}, M., {Lehner}, L., {Liebling}, S.~L., \&
  {Neilsen}, D. 2009, \prl, 103, 081101

\bibitem[{{Palenzuela} {et~al.}(2010{\natexlab{a}}){Palenzuela}, {Garrett},
  {Lehner}, \& {Liebling}}]{2010PhRvD..82d4045P}
{Palenzuela}, C., {Garrett}, T., {Lehner}, L., \& {Liebling}, S.~L.
  2010{\natexlab{a}}, \prd, 82, 044045

\bibitem[{{Palenzuela} {et~al.}(2010{\natexlab{b}}){Palenzuela}, {Lehner}, \&
  {Liebling}}]{2010Sci...329..927P}
{Palenzuela}, C., {Lehner}, L., \& {Liebling}, S.~L. 2010{\natexlab{b}},
  Science, 329, 927

\bibitem[{{Palenzuela} {et~al.}(2010{\natexlab{c}}){Palenzuela}, {Lehner}, \&
  {Yoshida}}]{2010PhRvD..81h4007P}
{Palenzuela}, C., {Lehner}, L., \& {Yoshida}, S. 2010{\natexlab{c}}, \prd, 81,
  084007

\bibitem[{Pollney {et~al.}(2007)Pollney, Reisswig, Rezzolla, Szil{\'a}gyi,
  Ansorg, Deris, Diener, Dorband, Koppitz, Nagar, \&
  Schnetter}]{Pollney:2007ss}
Pollney, D., Reisswig, C., Rezzolla, L., {et~al.} 2007, Phys. Rev. D, 76,
  124002

\bibitem[{{Rezzolla} {et~al.}(2011){Rezzolla}, {Giacomazzo}, {Baiotti},
  {Granot}, {Kouveliotou}, \& {Aloy}}]{2011ApJ...732L...6R}
{Rezzolla}, L., {Giacomazzo}, B., {Baiotti}, L., {et~al.} 2011, \apjl, 732, L6

\bibitem[{{Rossi} {et~al.}(2010){Rossi}, {Lodato}, {Armitage}, {Pringle}, \&
  {King}}]{2010MNRAS.401.2021R}
{Rossi}, E.~M., {Lodato}, G., {Armitage}, P.~J., {Pringle}, J.~E., \& {King},
  A.~R. 2010, \mnras, 401, 2021

\bibitem[{Schnetter {et~al.}(2004)Schnetter, Hawley, \&
  Hawke}]{Schnetter-etal-03b}
Schnetter, E., Hawley, S.~H., \& Hawke, I. 2004, Class. Quantum Grav., 21, 1465

\bibitem[Schnittman(2011)]{2011CQGra..28i4021S} Schnittman,
  J.~D.\ 2011, Class. Quantum Grav., 28, 094021

\bibitem[{{Schnittman} \& {Krolik}(2008)}]{2008ApJ...684..835S}
{Schnittman}, J.~D., \& {Krolik}, J.~H. 2008, \apj, 684, 835

\bibitem[{{Shapiro}(2010)}]{2010PhRvD..81b4019S}
{Shapiro}, S.~L. 2010, \prd, 81, 024019

\bibitem[{{Shields} \& {Bonning}(2008)}]{2008ApJ...682..758S}
{Shields}, G.~A., \& {Bonning}, E.~W. 2008, \apj, 682, 758

\bibitem[{{Tanaka} \& {Menou}(2010)}]{2010ApJ...714..404T}
{Tanaka}, T., \& {Menou}, K. 2010, \apj, 714, 404

\bibitem[{Thornburg(2004)}]{Thornburg2003:AH-finding}
Thornburg, J. 2004, Class. Quantum Grav., 21, 743

\bibitem[van Meter et al.(2006)]{2006PhRvD..73l4011V} van Meter,
  J.~R., Baker, J.~G., Koppitz, M., \& Choi, D.-I.\ 2006, \prd, 73,
  124011

\bibitem[{{van Meter} {et~al.}(2010){van Meter}, {Wise}, {Miller}, {Reynolds},
  {Centrella}, {Baker}, {Boggs}, {Kelly}, \&
  {McWilliams}}]{2010ApJ...711L..89V}
{van Meter}, J.~R., {Wise}, J.~H., {Miller}, M.~C., {et~al.} 2010, \apjl, 711,
  L89

\bibitem[{{Zanotti} {et~al.}(2010){Zanotti}, {Rezzolla}, {Del Zanna}, \&
  {Palenzuela}}]{2010arXiv1002.4185Z}
{Zanotti}, O., {Rezzolla}, L., {Del Zanna}, L., \& {Palenzuela}, C. 2010, \aap,
  523, A8

\end{thebibliography}

\end{document}